# Superconducting gap and its Little-Parks like oscillations with high-order harmonics in lithium intercalated *1T*-TiSe$_2$

Jia-Yi Ji[1#], Zongzheng Cao[1#], Yi Hu[1], Haoyang Wu[1], Heng Wang[1], Yuying Zhu[2], Haiwen Liu[3], Lexian Yang[1], Qi-Kun Xue[1,2,4]*, and Ding Zhang[1,2,5]*

[1]State Key Laboratory of Low Dimensional Quantum Physics and Department of Physics, Tsinghua University, Beijing 100084, China

[2]Beijing Academy of Quantum Information Sciences, Beijing 100193, China

[3]Center for Advanced Quantum Studies, Department of Physics, Beijing Normal University, Beijing 100875, China

[4]Southern University of Science and Technology, Shenzhen 518055, China

[5]Hefei National Laboratory, Hefei 230088, China

# These authors contributed equally.

**Corresponding author**

*Email: qkxue@mail.tsinghua.edu.cn

    dingzhang@mail.tsinghua.edu.cn

# Abstract

The superconducting phase of doped 1T-TiSe$_2$ is a fruitful playground for exploring exotic quantum phenomena such as the anomalous metal state and spontaneously formed superconducting network. Here, we address these emergent states by studying the superconducting gap of lithium intercalated TiSe$_2$—a fundamental quantity that has remained unexplored so far. We fabricate a device that combines solid-state lateral lithium intercalation, resistance measurements and tunneling spectroscopy. We successfully probe the superconducting gap of TiSe$_2$ and reveal that the gap closing temperature well exceeds the transition temperature ($T_c$) expected from the Bardeen-Cooper-Schrieffer theory, indicating pronounced superconducting fluctuations even in a bulk-like system. Moreover, the symmetric gap persists even in the anomalous metal state, demonstrating the particle-hole symmetry of this exotic phase directly from the density of states. Finally, the superconducting gap shows magneto-oscillations with higher harmonics, attesting to a rather regular structure of the intrinsic superconducting network.

Interplay of different orderings may give rise to emergent phenomena, such as the recently discovered pair density waves in strongly correlated systems [1–5]. Titanium diselenide is an interesting platform that hosts many collective quantum phenomena, including the charge density wave (CDW) [6–8], a gyrotropic electronic order [9], and superconductivity after electrostatic/chemical doping [10-15] or under high pressure [16]. In the superconducting state of doped TiSe$_2$, magneto-transport experiments further revealed resistance oscillations akin to the Little-Parks effect [13, 15]. Similar oscillations have also been observed in lightly doped Bi$_2$Sr$_2$CaCu$_2$O$_{8+x}$ [17]. It reflects a spontaneously emerged superconducting texture with a surprisingly large period (a few hundred nanometers). Of late, large domains of CDW have been identified in NbSe$_2$ [18], suggesting that their formation is energetically favorable. In bulk TiSe$_2$ under lithium intercalation, such a superconducting texture is responsible for the formation of yet another enigmatic state—the anomalous metal state [13], which otherwise appears only in a purely two-dimensional setting [19, 20].

So far, studies on TiSe$_2$ as a superconductor have largely focused on resistance characterizations [10-16], leaving many other fundamental properties less explored. Among them, the superconducting gap can provide pivotal information on the correlated nature of this system but remains scarcely addressed for doped TiSe$_2$. Determining the superconducting gap of TiSe$_2$ can shed light on the enigmatic gap opening observed by ultrafast optics [21]. Also, the ion gated TiSe$_2$ is a unique superconducting system that hosts anomalous metal state [19] even in a bulk-like situation [13]. Measuring the gap of the anomalous metal state has not been achieved so far, although it can help address directly the issue of particle-hole symmetry [22]. Moreover, the Little-Parks oscillations, as seen by resistance measurements, can in principle manifest themselves as gap oscillations too. For the classical Little-Parks effect, magnetic fluxes modulate the superconducting transition temperature $T_c$ of a ring with an amplitude of $\delta T_c$. Such a variation of $\delta T_c$ can result in a modulated superconducting gap $\Delta$: $\delta\Delta = \delta T_c \cdot \Delta/T_c$, with $\Delta/T_c$ being the gap-to-$T_c$ ratio. Such a property has not been clearly identified [23].

Here, we carry out tunneling spectroscopy on lithium intercalated 1T-TiSe$_2$ (Li$_x$TiSe$_2$). We continuously tune the doping concentration in TiSe$_2$ via the lateral intercalation of lithium from an ion backgate. We combine this solid-state gating with the on-chip tunnel junction, allowing us to resolve the superconducting gap of the doped TiSe$_2$. Interestingly, we obtain a gap closing temperature that is prominently larger than that from the BCS gap-to-$T_c$ ratio. In the anomalous state, we observe that the gap remains to be symmetric. Furthermore, the gap shows prominent oscillations with a perpendicular magnetic field. It not only confirms the expected behavior of Little-Parks like oscillations but also reveals higher harmonics, indicating that the periodic superconducting texture is quite regularly shaped.

We start by introducing the measurement setup [Fig. 1(a)]. It combines three functions: (1) lateral lithium intercalation via a solid ion conductor [24, 25]; (2) resistance characterization; (3) tunneling spectroscopy. An optical image of the device is provided in Fig. S1. We employ a lateral intercalation scheme for improving the sample homogeneity [26-29]. By applying a backgate voltage $V_g$, lithium ions first intercalate into one end of TiSe$_2$ that is directly on the solid ion conductor [left region in Fig. 1(a)]. Afterwards, they migrate (horizontal arrows) via the interlamellar gaps of TiSe$_2$ [28]. In the laterally intercalated region of TiSe$_2$ on the AlO$_x$ buffer, we carry out either resistance (highlighted by a purple rectangle) measurements or the tunneling (black rectangle) spectroscopy (details in the Methods section). They are typically done at low temperatures such that lithium ions are completely immobile.

Figure 1(b) shows the temperature dependent resistances of Li$_x$TiSe$_2$. Traces are from the same sample but correspond to four different gated states. The carrier density of each state is determined by the Hall effect measurement at low

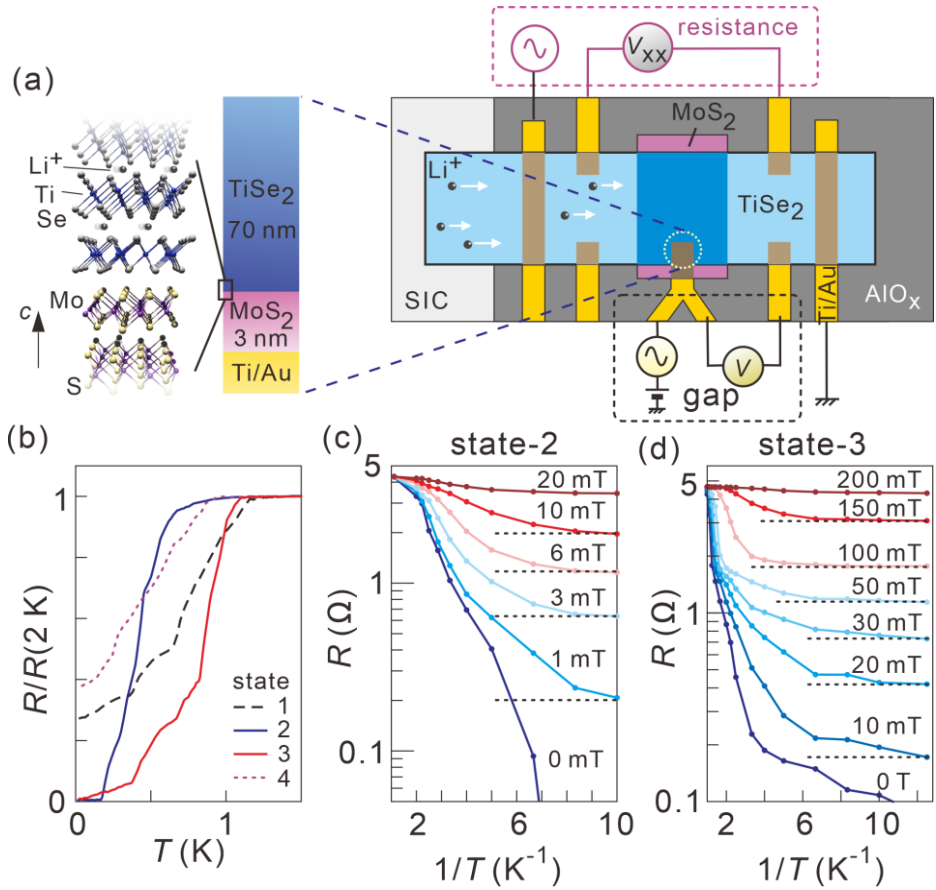

**Figure 1** *(a) Schematic drawing of the experimental setup. Left panel illustrates the crystalline structure of lithium intercalated 1T-TiSe$_2$ on top of 2H-MoS$_2$ at the interface. Middle panel sketches the cross-sectional view of the tunnel junction with thicknesses of the constituent materials. Right panel shows the device that combines ion gating, resistance measurements (module marked out by the purple rectangle) and tunneling spectroscopy (module marked out by the black rectangle). White arrows indicate the intercalation of lithium ions. (b) Temperature dependent resistance of the TiSe$_2$ flake after lithium intercalation. Here the resistance is normalized by the value at 2 K. Numbers mark the consecutive gating sequence. The carrier densities of states 1 to 4 are $(5.5, 6.9, 13, 38) \times 10^{15}$ cm$^{-2}$. (c) and (d) Arrhenius plots of resistances at selected magnetic fields for state-2 and state-3. Dashed horizontal lines are guide to the eye.*

temperatures (details in Fig. S2). The doping level with the highest $T_c$ achieved in this sample is at around $n = 1.3 \times 10^{16}$ cm$^{-2}$ [state-3, red curve in Fig. 1(b)]. The corresponding $T_c$, if taking the point where the resistance drops to half of the normal state resistance, is 0.85 K. The superconducting transition for state-3 is broader than that of state-2. It suggests stronger superconducting fluctuations. In fact, for both

state-2 and state-3, we observe anomalous metal states as shown in Fig. 1(c)(d). Under a finite magnetic field, the resistance of Li$_x$TiSe$_2$ shows a plateau at low temperatures, as indicated by the horizontal lines. We remark that the present study is carried out in a cryogenic system with a series of low temperature filters: thermocoax cables [30] from the ports at room temperature to those on the mixing chamber; copper powder filters [31] and RC filters (cutoff at 10 kHz) on the mixing chamber. The persistence of saturated resistance at low temperatures after all the filtering strongly suggests an intrinsic anomalous metal state [32–34].

We now carry out tunneling spectroscopy. The tunnel junction consists of a Ti/Au metal electrode, a thin barrier of MoS$_2$, and lithium intercalated TiSe$_2$ [left panels of Fig. 1(a)], i.e., a normal metal-insulator-superconductor (NIS) junction. We show in Fig. S3 that lithium ions cannot intercalate into MoS$_2$ layer in this scheme. Away from the tunnel junction, the metal electrode bifurcates. This design allows for a four-terminal measurement [black rectangle in Fig. 1(a)]. We point out that the tunneling spectroscopy measures a much more local area of the intercalated sample. Figure 2(a) and (b) collect the tunneling spectra at different temperatures under zero magnetic field for state-2 and state-3, respectively. The spectra at low temperatures show typical features of a superconducting gap—a previously unaddressed aspect of TiSe$_2$ based superconductor: two superconducting coherent peaks and suppressed conductance around zero bias. With increasing temperature, the coherent peaks gradually disappear and the valley around zero bias gets filled up.

We further fit the tunneling spectra for state-2 and state-3 by using the Blonder-Tinkham-Klapwijk (BTK) formula [35] (details in Fig. S5), assuming *s*-wave pairing [36, 37]. Figure 2(c) and (d) show the superconducting gap values at different temperatures. In the zero-temperature limit, Δ= 33 and 76 μeV for state-2 and state-3, respectively. Based on the BCS gap-to-$T_c$ ratio ($2\Delta/T_g = 3.53$), we estimate the superconducting transition temperature $T_c^{BCS}$ to be 0.22 and 0.50 K for the two states (marked by dotted lines in Fig. 2(c)(d)). They are substantially lower than the gap closing temperatures $T_g$: 0.42 and 0.76 K (indicated by arrows in Fig. 2(b)). From another perspective, $2\Delta/T_g$ is about 1.82 and 2.31, which are substantially smaller than the BCS ratio. This is unusual for TiSe$_2$ with a well-defined electron pocket [38]. In general, it is rare for superconductors to host $2\Delta/T_g$ smaller than 3.53. For instance, a recent scanning tunneling microscopy study reveals that Cu$_{0.06}$TiSe$_2$ possesses $2\Delta/T_g = 4.3$ and the temperature dependence of the gap follows nicely the BCS theory [39]. We therefore conclude that lithium intercalation acts differently than Cu doping.

Notably, the large discrepancy between $T_c^{BCS}$ and $T_g$ with $T_c^{BCS} < T_g$ was previously reported only in 2D superconductors. Such an unusual phenomenon was attributed to either strong phase fluctuations [40, 41] or the crossover from BCS Cooper pairing to a Bose Einstein Condensate (BEC), i.e., BCS-BEC crossover [26, 42]. For the latter case, $\Delta$ should be comparable to the Fermi level $E_F$. By contrast, we estimate $E_F$ of Li$_x$TiSe$_2$ to be 100-200 meV for state-2 and state-3, such that $\Delta/E_F \sim 3 \times 10^{-4}$. The BCS-BEC crossover is therefore unlikely to account for $T_c^{BCS} < T_g$ in Li$_x$TiSe$_2$. For the other mechanism of superconducting fluctuations, we note that our sample has a thickness of 70 nm, which can be treated as a bulk system. It indicates that Li$_x$TiSe$_2$ is quite unique that it hosts comparable superconducting fluctuations to those in the 2D limit. In fact, the anomalous metal state we observed are caused by such strong superconducting fluctuations. Additionally, a recent ultrafast optical measurement observed a superconductor-like gap in TiSe$_2$: 2 meV at 4 K [21], which is 1-2 orders of magnitude larger than the superconducting gap we obtain in Li$_x$TiSe$_2$. This large discrepancy is distinctly different from the case of K$_3$C$_{60}$ [43], in which the light-induced high-temperature superconductivity and equilibrium superconductivity share the same gap. It suggests that the light-induced gap in TiSe$_2$ may have a different origin than that in K$_3$C$_{60}$.

The tunneling spectroscopy allows us to address another critical issue: the particle-hole symmetry in the anomalous metal state. So far, experiments demonstrating this unique property has been limited to Hall effect measurements. Zero Hall resistance in the regime of anomalous metal state indicates particle-hole symmetry [22, 33, 44]. However, this method requires a careful distinction between a small but non-zero Hall signal and the truly zero one. In contrast, our setup unveils the particle-hole symmetry directly from the density of states. Figure 2(e) presents typical tunneling spectra at 10 mT, 20 mT, 30 mT at 0.1 K. For this chosen set of magnetic field values and temperature point, resistance measurements confirm that the sample is in the anomalous metal state [Fig. 1(d)]. As the magnetic field increases, the superconducting gap becomes gradually suppressed but remains symmetric.

For a quantitative analysis, we evaluate the height of the superconducting coherent peaks [indicated in Fig. 2(e)] and the suppressed spectral weights inside the gap, in comparison to the spectrum at 1.4 K, for $V > 0$ and $V < 0$ separately [$S_+$ and $S_-$ marked in Fig. 2(e)]. Figure 2(f)(g) demonstrate that these extracted quantities remain comparable in the positive and negative sections, irrespective of the applied magnetic fields. This symmetry around zero bias directly reflects the particle-hole symmetry of the anomalous metal state [45]. At 20 mT and 30 mT, the superconducting coherence peaks are strongly suppressed but the dip in the density of states around zero bias persists. It indicates that the quantum coherence is suppressed in the anomalous metal state. This behavior confirms the theoretical expectation that the anomalous metal state hosts incoherent Cooper pairs, which account for the finite resistance as $T \to 0$ [20, 46]. The method demonstrated here could be employed to study the anomalous metal state, of either intrinsic or extrinsic origin, in other material systems such as NbSe$_2$ [44], PdTe$_2$ [34], etc.

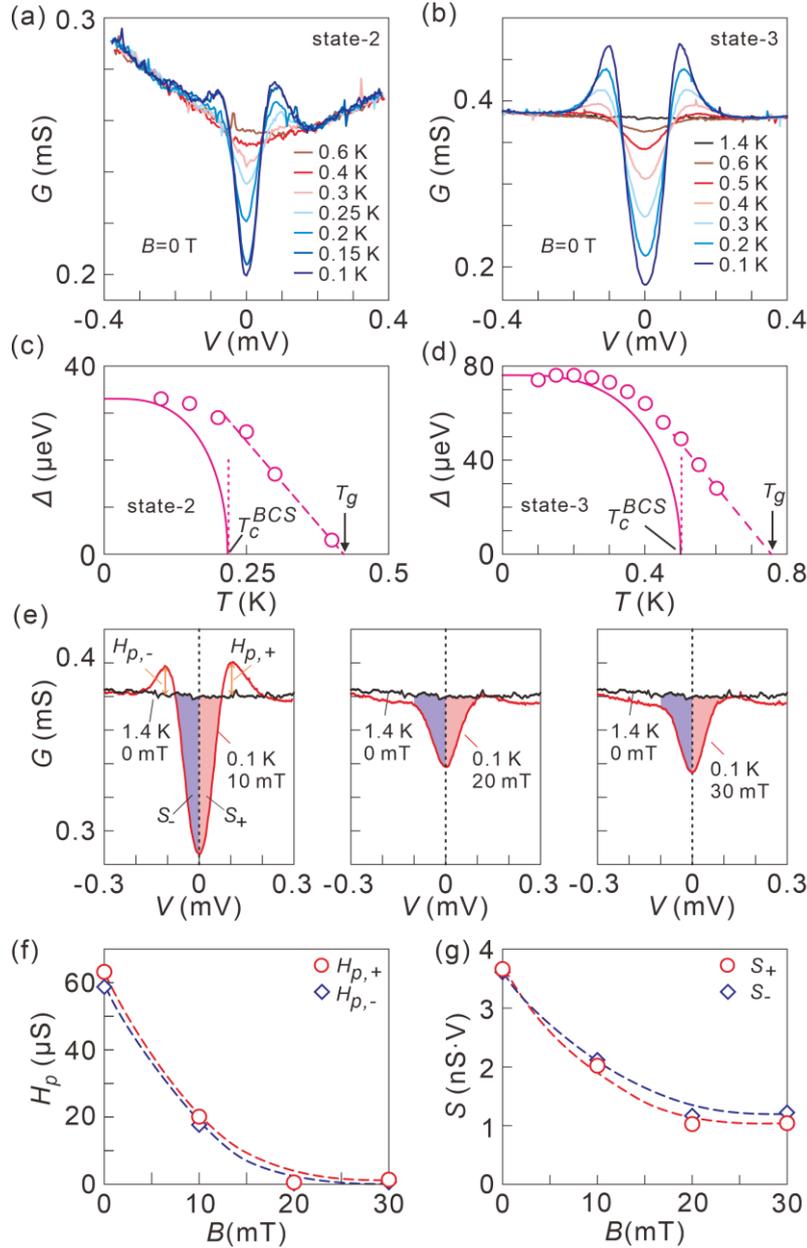

*Figure 2* (a)(b) Tunneling spectra of state-2 and state-3 under zero magnetic field at various temperatures. (c)(d) Temperature dependences of the superconducting gap for state-2 and state-3. Arrows indicate the gap closing temperature ($T_g$). Solid curves are the theoretically expected temperature dependences based on the BCS model. Dotted lines mark the superconducting transition temperature based on the BCS gap-to-$T_c$ ratio of 3.53 ($T_c^{BCS}$). (e) Red: tunneling spectra at selected magnetic fields in the anomalous metal state at 0.1 K for state-3. Black: zero-field tunneling spectrum at 1.4 K. Shaded regions highlight the suppressed density of states within the superconducting gap for $V>0$ or $V<0$. (f) Heights of superconducting coherent peaks as a function of magnetic field. (g) Areas of the superconducting gap with $V>0$ and $V<0$ as a function of magnetic field. They are extracted from the data shown in panel (b) and (e).

We now present the third major finding of this work—the magneto-oscillations of the superconducting gap in Li$_x$TiSe$_2$. Figure 3(a)(b) show the color-coded tunneling spectra as a function of bias voltage and magnetic field for the sample at state-3. They are obtained by sweeping the magnetic field in the positive (a) and negative (b) directions. In order to suppress spurious effects, we sweep the magnetic field multiple times with different but low sweep rates (details in Methods). We then average over the data obtained in the same sweeping direction. The results in the positive and negative sweeps are antisymmetric (discussed in Fig. S7). Both data sets show that the blue region expands and shrinks in an alternating manner along the magnetic field axis. Correspondingly, the red regions at around $\pm 0.1$ mV—the superconducting coherence peaks—oscillate. It demonstrates that the oscillation occurs for the complete superconducting gap. The subtraction between the data in the two panels produces results in Fig. 3(c). It highlights the regular oscillations as alternating regions with positive or negative values as a function of magnetic field. Fig. 3(d) collects the zero-bias tunneling conductance $G_V = 0$. The hysteresis between the positive and negative sweeps indicates trapping of vortices [47, 48]. The pinning centers can arise from both structural defects and inhomogeneous lithium intercalation. We comment that the oscillatory features occur at the same perpendicular magnetic field, irrespective of the varying in-plane magnetic field (discussed in Fig. S8). It suggests a two-dimensional electronic structure that is responsible for the oscillations in the sample [13]. Furthermore, oscillations vary with doping. For state-2 (data shown in Fig. S9), gap oscillations are squeezed to a narrower range of magnetic fields, due to the weakened superconductivity.

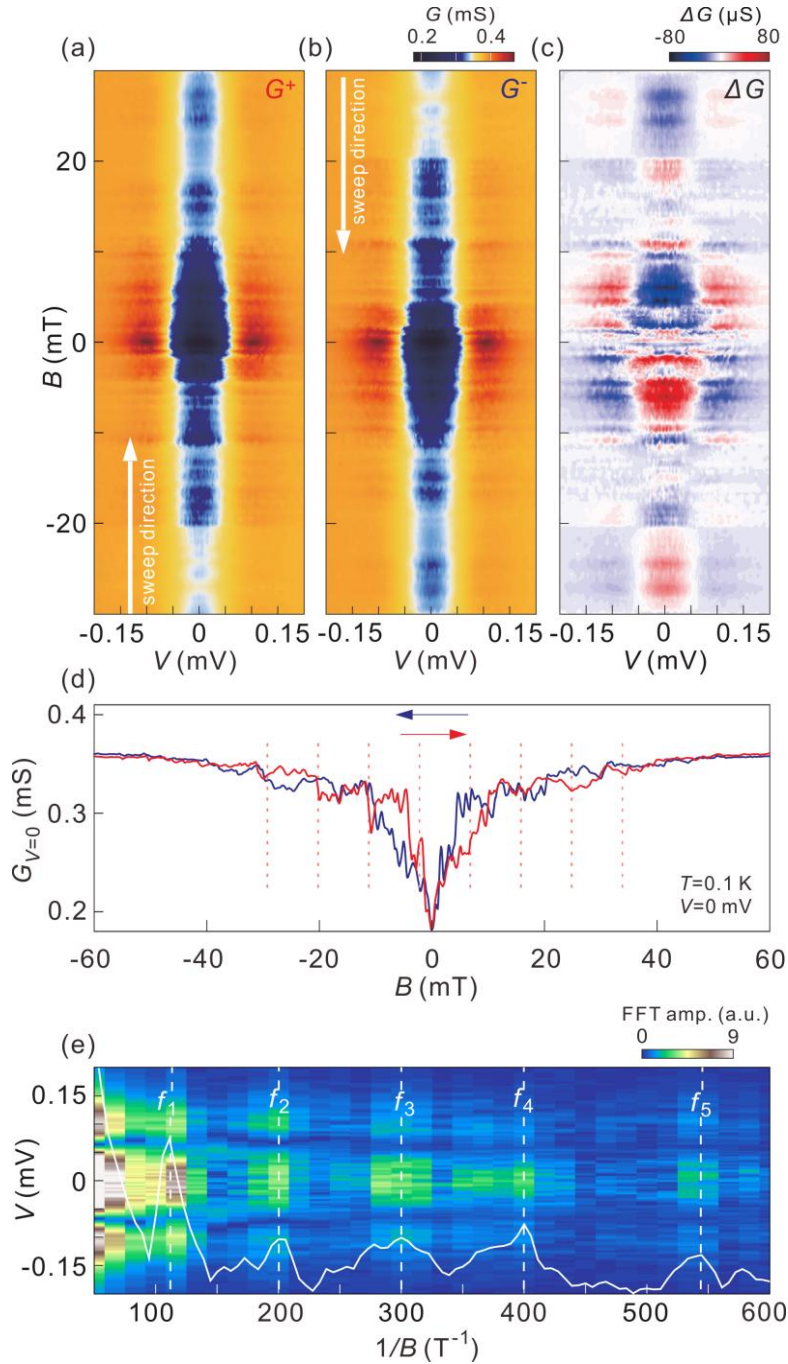

***Figure 3*** *(a)(b) Color-coded tunneling conductance as a function of bias voltage and magnetic field in the positive and negative sweep directions. The temperature is 0.1 K. (c) Subtraction of the tunneling conductance in positive by negative sweep as a function of bias voltage and magnetic field. (d) Zero bias tunneling conductance as a function of magnetic field for $Li_xTiSe_2$ at state-3. Red and blue curves are from positive and negative sweeps. Dotted lines are equidistant with a separation of 9.01 mT—a period determined by FFT as will be explained in (e). (e) Color-coded FFT of the data in (a). White solid line is the FFT of the data in the positive sweep of (d). Dashed lines indicate the five prominent peaks.*

Figure 3(e) shows the fast Fourier transforms (FFT) of the tunneling conductance in the positive sweep. It is plotted as a function of magnetic field and bias voltages. The white curve is the FFT of the trace in the positive sweep in Fig. 3(d). For both the color map and the white curve, we can identify five peaks, marked as $f_1$ to $f_5$. Importantly, these FFT peaks cover the complete voltage range of the superconducting gap. From the frequency value $f_1 = 111$ T$^{-1}$, we obtain the oscillation period: $\Delta B = 1/f_1 = 9.01$ mT (highlighted as equidistant lines in Fig. 3(d)). Such a periodic modulation under a magnetic field reflects entrance or exit of fluxes into a certain area encircled by the superconducting loop. The size of such an area $S_1$ can be estimated by using the equation:

$$S_1 \Delta B = h/2e. \tag{1}$$

Here the right-hand side of Eq. (1) is the flux quantum. By inserting the extracted value of $\Delta B$ from experiments, we estimate $S_1$ to be 0.23 μm², which corresponds to a lateral size of about 730 nm if assuming a triangular lattice. The estimated area is one order of magnitude smaller than the area of our tunnel junction 3.4 μm² (Fig. S1). For state-2, FFT gives rise to a pronounced peak at 327 T$^{-1}$ such that the corresponding area for the fluxoid quantization is about 0.68 μm², still much smaller than the junction. It indicates that there exists a network of multiple superconducting and non-superconducting regions in our sample.

In Fig. 4(a), we plot the FFT peak positions as a function of the assigned sequence numbers from 1 to 5 for $f_1$ to $f_5$. Interestingly, the first four peaks fall nicely on the same trend that crosses the origin (solid line). The fifth peak also stays close to the fitted line. This scaling suggests that $f_1$ is the base frequency while others are higher harmonics: $f_m = mf_1, \ m = 2, 3, 4$. The higher harmonics indicate that Eq. (1) should be generalized as:

$$S_m = mf_1 h/(2ne), \tag{2}$$

where $S_m$ is the area of the superconducting loop for the $m$-th harmonic and $n$ is an integer. There exist two scenarios to account for the high harmonics. First, if $S_m$ is required to be the same as $S_1$, then $n$ increases together with $m$ such that $m/n =$

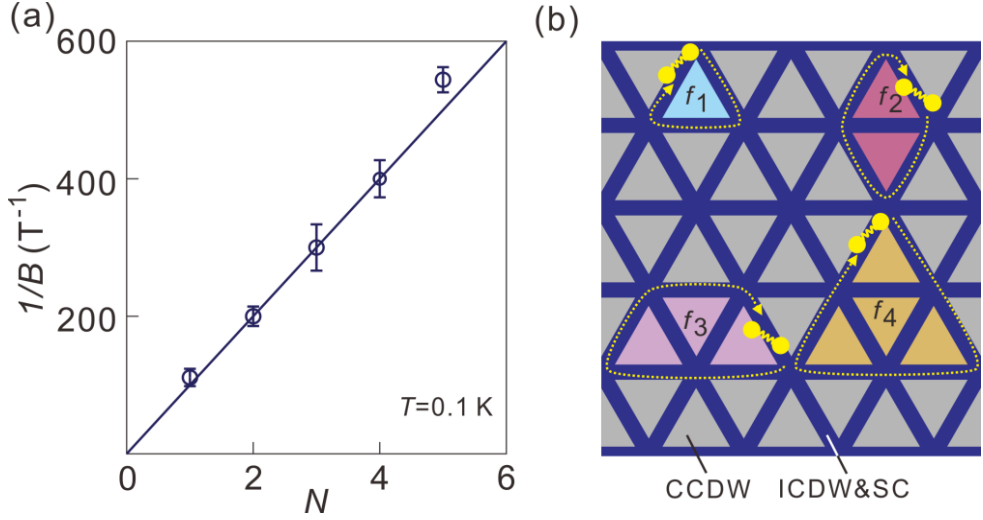

***Figure 4*** *(a) FFT peak positions as a function of the assigned sequence (from one to five). Error bars are the full width at half-maximum of FFT peaks. (b) Schematic drawing of the superconducting texture. Gray triangles indicate the non-superconducting regions consisted of CCDW domains. Dark blue stripes indicate ICDW domain boundaries that are superconducting at low temperatures. Triangles highlighted by brighter colors illustrate the circulated area for the Cooper pairs that can generate the base frequency and higher harmonics.*

1. For $n = 2, 3, 4$, it indicates that multiple Cooper pairs participate together in circling around the non-superconducting area of $S_1$. This is similar to the recent observation of charge-4e and charge-6e flux quantization in CsV$_3$Sb$_5$ [49]. Despite the fact that TiSe$_2$ hosts chiral ordering, we argue that higher-charge superconductivity is unlikely because: (1) charge-4*e* and charge-6*e* in CsV$_3$Sb$_5$ were observed at temperatures around $T_c$. By contrast, we observe the higher harmonics at $T \ll T_c$; (2) For the harmonic at $f_5$, applying the scenario of higher charge superconductivity to explain our results would require charge-10*e* pairing.

We thus explore the second scenario for the higher harmonics. Namely, $n$ stays to be 1 and $S_m$ becomes multiples of $S_1$: several superconducting loops combine together to form a larger loop for the charge-2*e* Cooper pairs to travel around. Such a scenario is schematically shown in Fig. 4(b). The frequencies $f_2$ to $f_4$ correspond to the enlarged loop around two to four non-superconducting domains. In practice, the

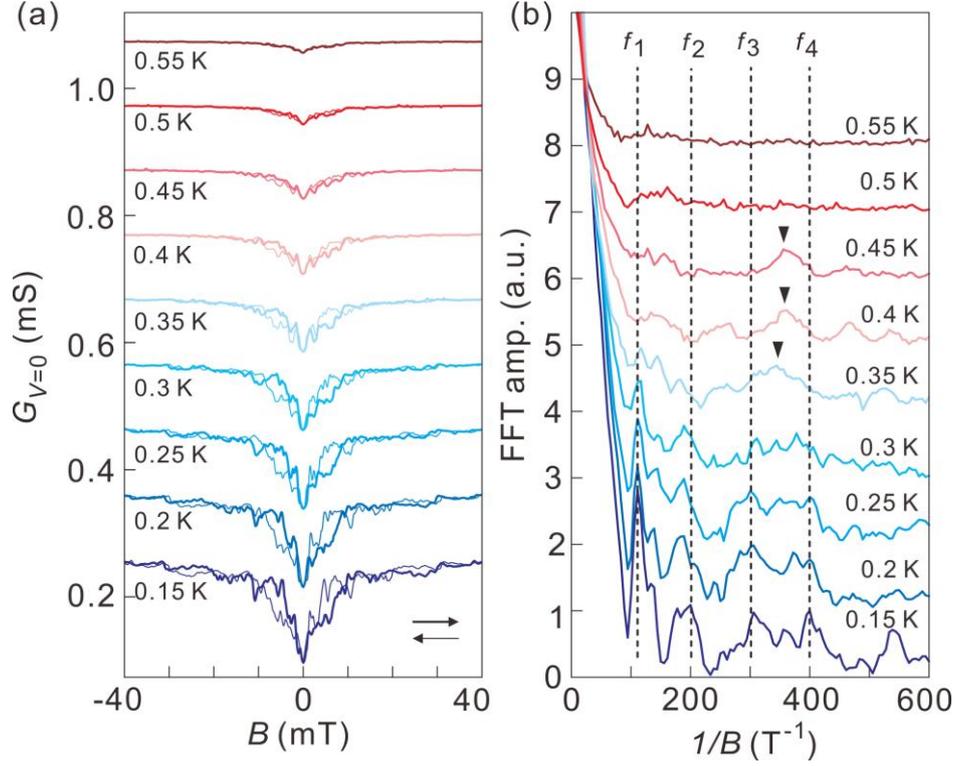

**Figure 5** (a) Zero bias tunneling conductance $G_{V=0}$ as a function of magnetic field at different temperatures for Li$_x$TiSe$_2$ at state-3. Thick (thin) curves are obtained in the positive (negative) sweep directions. Curves are vertically offset for clarity. (b) Fast Fourier transforms (FFT) of $G_{V=0}$ in the positive sweep at different temperatures. Curves are vertically offset for clarity. Dashed lines mark the peak positions at 0.1 K.

shape of the domain may deviate from the ideally periodic case [18]. This deviation may account for the anharmonic behavior at the high frequency of $f_4$. We also comment that the lack of higher harmonics in previous experiments on ion gated TiSe$_2$ may stem from the fact that resistance measurements average out the signal over a large area, whereas tunneling spectroscopy is more local.

We further investigate the temperature dependence of gap oscillations. Figure 5(a) shows the zero-bias tunneling conductance as a function of magnetic field at elevated temperatures. Figure 5(b) shows the corresponding FFT of each trace in the positive sweep. Clearly, the peaks observed at 0.1 K persist to higher temperatures of about 0.3 K. There, we can still identify peaks at $f_1$ and $f_2$ but the higher harmonics become hardly discernible. At $T > 0.3$ K, we observe peaks at distinctly different positions. Especially, the peak at about 350 T$^{-1}$ stays from 0.35 K to 0.45 K, as marked by the black triangles. It corresponds to $S = 0.72$ μm$^2$, which is larger than $S = 0.23$ μm$^2$ for $f_1$ at 0.1 K. This variation may reflect that multiple non-superconducting domain merge together with increasing temperature. Such a temperature dependence is opposite to that observed in the resistance measurements previously [13]. This contrast suggests that the Little-Parks like oscillations may have strong doping and spatial dependences.

Our tunneling study complements to the previous resistance measurements on establishing the spontaneously formed superconducting network in TiSe$_2$. Such a periodic structure likely consists of commensurate CDW domains with incommensurate CDW domain walls [13, 15] [Fig. 4(b)]. Indeed, the natural tendency toward incommensuration of the otherwise commensurate CDW in TiSe$_2$ has been confirmed by X-ray diffraction as well as scanning tunneling microscopy on TiSe$_2$ under pressure [50] or Cu-doping [11, 12]. Importantly, the ICDW coexists with superconductivity in the same pressure or doping range of the phase diagram. The enhanced density of states along the ICDW regions favors superconductivity. By contrast, the commensurate CDW domains may stay non-superconducting. The dichotomous behaviors of ICDW and CCDW give rise to a superconducting network, which in turn causes Little-Parks oscillations and strong superconducting fluctuations. We note that the exact structure of CCDW domains may be dopant dependent. Li$_x$TiSe$_2$ hosts a nearly constant onset temperature for CDW (Fig. S10), similar to protonated TiSe$_2$ [8, 14] but sharply different from the decreasing $T_{CDW}$ with doping in Cu$_x$TiSe$_2$. In addition, we note that there exists another type of domain wall between the chiral CDW domains in TiSe$_2$. This type of domain wall may not be relevant because it contributes no additional electronic states [51].

In summary, we unveil the superconducting gap of Li$_x$TiSe$_2$. We observe strongly enhanced superconducting fluctuations in bulk-like samples, manifesting as a substantially higher gap closing temperature than $T_c$ predicted by BCS theory. The tunneling spectra are symmetric in the anomalous metal state, demonstrating the particle-hole symmetry. Furthermore, the gap oscillates with a perpendicular magnetic field and the oscillation contains higher harmonics. Our work opens a distinct pathway to address emergent phenomena under *in-situ* ion gating.

**Methods:**

The solid ion conductor (SIC) [25] has a chemical formula of $Li_2O$-$Al_2O_3$-$SiO_2$-$P_2O_5$-$TiO_2$-$GeO_2$. To fabricate the device, we first deposit a square area ($100 \times 100$ μm$^2$) of AlO$_x$ film (15 nm) with a rate of 0.25 nm/min on the solid ion conductor by using standard electron-beam lithography and atomic layer deposition (ALD). Our previous experiments showed that a TiSe$_2$ flake solely on AlO$_x$ cannot be intercalated by lithium ions from the solid ion conductor [28]. Therefore, the AlO$_x$ film is an effective blocking layer against lithium intercalation. The substrate was then patterned with Ti/Au (10 nm/30 nm) electrodes and loaded into the argon glove box ($H_2O$<0.1 p.p.m. and $O_2$<0.1 p.p.m.).

In the glovebox, we dry transferred the mechanically exfoliated flakes of MoS$_2$ and TiSe$_2$ by using the gel-film [28][29]. The MoS$_2$ flake was about 3 nm and covered the electrodes for tunneling measurements on top of the AlO$_x$ buffer. It served as the tunnel barrier with a resistance of about 3 kΩ at 0.1 K. The TiSe$_2$ flake had a thickness of 70 nm and covered both the AlO$_x$ buffered region and the exposed surface of the solid ion conductor. After the stacking process, the sample was taken out for measurements.

Ion back-gating and electrical measurements were carried out in a dilution refrigerator (Bluefors SD) equipped with a home-built piezo-rotator (Attocube ANRv220). The rotator has a calibrated angular precision of 0.006 degree. The dilution refrigerator has 24 thermo-coax cables [30] from the room-temperature side to the mixing chamber. These cables are then connected to a copper powder filter [31] and a RC filter, both on the mixing chamber plate. As shown in Fig. S11, the performance of this combination of thermos-coax, copper power/RC filters is on the same level as those used for suppressing the extrinsic anomalous metal state [32]. The lowest electron temperature of this system reaches about 80 mK, which is calibrated by measuring the thermally activated behavior of fractional quantum Hall states (FQHS) in a GaAs two-dimensional electron gas (2DEG) on the same rotator (Fig. S12).

Electrical measurements were performed with the standard lock-in technique. The excitation current was chosen to be 1 µA (13.333 Hz). The tunneling conductance was measured by using heterodyne detection with an ac current of 4 nA at 13.333 Hz in a four-terminal configuration. For measuring the tunneling conductance as a function of magnetic field, we swept four times back and forth with the following sweep rates: 5.0 mT/min, 5.3 mT/min, 5.7 mT/min, 6.0 mT/min. We then averaged the signal in the same sweep direction. In order to remove the small remanent field of the solenoid magnet, we chose the global minimum of the conductance trace as the point where the magnetic field is zero (details in Fig. S6). The angular dependent study was carried out by rotating the sample in situ. We calibrate the rotation angle by measuring the Hall effect of the same sample at 0.1 K (details in Fig. S8).

## ASSOCIATED CONTENT

**Supporting information**

This material is available free of charge via the internet at http://pubs.acs.org.

Optical and scanning electron microscopy images of the device; Hall effect measurement for lithium intercalated $TiSe_2$; Ion gating of a molybdenum disulfide flake; Additional data on magnetoresistance oscillation of $Li_xTiSe_2$; Supporting data on the tunneling measurement.

## AUTHOR INFORMATION

**Corresponding Authors**

*E-mail: qkxue@mail.tsinghua.edu.cn;

 dingzhang@tsinghua.edu.cn

**Author contributions**

J.-Y. J. and Z. C. contributed equally. J.-Y. J. fabricated the samples. J.-Y. J. and Z. C. carried out transport measurements with the assistance of Y. H., H. Wu, H. Wang, Y. Z. J.-Y. J., Z. C., and D.Z. analyzed the data. D.Z., J.-Y. J. wrote the paper with the input from Z.C., L. Y., H. L., and Q.-K. X.

**Notes**

The authors declare no competing financial interest.

**Acknowledgements**

This work is financially supported by the Ministry of Science and Technology of China (2022YFA1403103); National Natural Science Foundation of China (Grants No. 12274249, No. 52388201).